\documentstyle{lamuphys}
\input psfig.tex
\makeatletter
\let\chapter\hid@chapter
\makeatother

\newcommand{\dalpha}{\dot{\alpha}}
\newcommand{\dbeta}{\dot{\beta}}

\newcommand{\bD}{\bar{D}}
\newcommand{\cD}{{\cal D}}
\newcommand{\cbD}{\bar{{\cal D}}}

\newcommand{\bQ}{\bar{Q}}
\newcommand{\blambda}{\bar{\lambda}}

\def\be{\begin{equation}}
\def\ee{\end{equation}}
\def\bea{\begin{eqnarray}}
\def\eea{\end{eqnarray}}

\def\cropen#1{\crcr\noalign{\vskip #1}}
\def\crr{\cropen{1\jot }}
\def\Red{}
\def\Black{}

\def\cA{{\cal A}}

\def\plushc{\ +\ \mbox{h.c.}}

\newcommand{\cL}{{\cal L}}
\renewcommand{\cD}{{\cal D}}
\newcommand{\cF}{{\cal F}}

\def\d{{\rm d}}

\catcode`@=11
\def\citenum#1{\csname b@#1\endcsname}
\catcode`@=12

\newcommand{\nn}{\nonumber}

\newcommand{\AmS}{{\protect\the\textfont2
 A\kern-.1667em\lower.5ex\hbox{M}\kern-.125emS}}

\begin{document}
\pagenumbering{arabic}
\title{Linear and Nonlinear Supersymmetries}

\author{Jonathan Bagger and Alexander Galperin}

\institute{Department of Physics and Astronomy,
 Johns Hopkins University \\
 3400 N.  Charles Street,
 Baltimore, MD 21218, USA}

\maketitle

\begin{abstract}
In this talk we use nonlinear realizations to study the
spontaneous partial breaking of rigid and local supersymmetry.
\end{abstract}
\section{Introduction}

The winter of 1996 was a hard one for physics, bringing
the untimely deaths of Professors Dmitrij Vasilievich
Volkov and Victor Isaacovitch Ogievetsky.  At this
symposium it seems appropriate to celebrate the memories
of both men, whose scientific achievements were so
closely aligned, and whose inspiring presence is already
acutely missed by their many friends and colleagues
across the world.

In this talk we will discuss a subject close to their
hearts:  supersymmetry and its nonlinear
realizations.  In particular, we will consider the partial
breaking of extended supersymmetry.  For simplicity, we
will restrict our attention to the case $N=2 \rightarrow
N=1$, but many of our results can be readily extended to
the case of higher supersymmetries, spontaneously broken
to $N=1$.  It is a fitting memorial to see many of
the ideas pioneered by Professors Volkov and Ogievetsky
come into play.

The partial breaking of supersymmetry is of crucial
importance to understanding the relation of theory
to experiment.  As theorists, we know in our bones
that there is an ultimate theory, perhaps M theory,
that exists at high energies.  However, this theory
is far removed from the physical world.  To connect
the two, we must integrate out the degrees of freedom
associated with the high energies and construct a
nonrenormalizable, effective field theory.  This
effective field theory should contain only those
degrees of freedom that are relevant for physics
in the world today.

Indeed, it is the point of view that underlies the
effective field theory approach to pion dynamics.
Below the scale of chiral symmetry breaking, we
know that the interactions of pions and hadrons are
governed by an effective field theory in which the
unbroken isospin symmetry is realized linearly, but
the spontaneously broken chiral symmetry is realized
nonlinearly.  The nonlinear symmetry is all that
remains of the chiral symmetry below the scale
where it is broken.

For the case at hand, we wish to construct a Lagrangian
with two supersymmetries.  The first supersymmetry,
that of $N=1$, is realized linearly, so it can be
represented in terms of superfields.  The second
supersymmetry, $N=2$, is realized nonlinearly on the
superfields.  In this way we can construct an effective
field theory of partial supersymmetry breaking.  This
theory is valid up to the scale where the second
supersymmetry is spontaneously broken.

At first glance, it might seem impossible to partially
break $N=2$ to $N=1$.  The argument runs as follows.
Start with the $N=2$ supersymmetry algebra,
\bea
\{ Q_\alpha,\,\bar
Q_{\dot\alpha} \} &\ =\ & 2\, \sigma^m_{
\alpha\dot\alpha}\,P_m \nonumber \\
\{ S_\alpha,\,\bar
S_{\dot\alpha} \} &=& 2\, \sigma^m_{
\alpha\dot\alpha}\,P_m\ ,
\label{susyalg}
\eea
where $Q_\alpha$ and its conjugate $\bar Q_{\dot\alpha}$
denote the first, unbroken supersymmetry, and $S_\alpha$,
$\bar S_{\dot\alpha}$ the second.  Suppose that one
supersymmetry is not broken, so
\be
Q\, |0\rangle \ =\ \bar Q\, |0\rangle\ =\ 0\ .
\ee
Because of the supersymmetry algebra, this implies that
the Hamiltonian also annihilates the vacuum,
\be
H\, |0\rangle \ =\ 0\ .
\ee
Then, according to the supersymmetry algebra,
\be
(\bar S S + S \bar S)\, |0\rangle\ =\ 0\ .
\label{SSbar}
\ee
The final step is to peel apart this relation and conclude
that
\be
S\, |0\rangle\ =\ \bar S\, |0\rangle\ =\ 0\ .
\ee
From this line of reasoning, one might think that partial
breaking is impossible.

Fortunately, this argument has two significant loopholes.
The first is that, technically-speaking, spontaneously-broken
charges do not exist.  Indeed, in a spontaneously broken
theory, one only has the right to consider the algebra of the
{\it currents.}  For the case at hand, the current algebra
can be modified as follows,
\bea
\{ \bar Q_{\dot\alpha},\
J^1_{\alpha m}
\} &\ =\ & 2 \,\sigma^n_{\alpha\dot\alpha}\,
T_{mn}\nonumber \\
\{ \bar S_{\dot\alpha},\ J^2_{\alpha m}
\} &=& 2\, \sigma^n_{\alpha\dot\alpha}\,
(v^4 \eta_{mn}+ T_{mn})\ ,
\label{current algebra}
\eea
where the $J^i_{\alpha m}$ ($i = 1,2$) are the supercurrents
and $T_{mn}$ is the stress-energy tensor.  Note that Lorentz
invariance does not force the right-hand sides of the commutators
to be the same.  If there were no first supersymmetry, the $v^4$
term in the second commutator could be absorbed in $T_{mn}$; it
would represent the scale of the supersymmetry breaking.  Now,
however, the first supersymmetry can be said to {\it define}
the stress-energy tensor, in which case there is an extra term
in the second commutator.  This discrepancy prevents the
current algebra from being integrated into a charge algebra,
and the no-go theorem is avoided.

The second loophole involves the last step of the theorem.
Even if the supercharges were to exist, it is only possible
to extract (5) from (4) if the Hilbert space is positive
definite.  In covariantly-quantized supergravity theories,
this is not the case: the gravitino $\psi_{m\alpha}$ is a
gauge field with negative-norm components.

There are, by now, many examples of partial supersymmetry
breaking which exploit the first loophole.  The first was
given by Hughes, Liu and Polchinski (1986) who showed
that supersymmetry is partially broken on the world volume
of an $N=1$ supersymmetric three-brane traveling in
six-dimensional superspace.  Since then there has been an
explosion of interest in membranes, so the number of
examples has grown substantially.  [For another
type of example, see Antoniadis, Partouche and Taylor
(1996).]

The membrane approach leaves many open questions.  For
example, we would like to know all possible field-theoretic
realizations of partial supersymmetry breaking, even those
that do not originate with branes.  We would also like to
know whether the $N=2$ supersymmetry gives rise to any
restrictions on matter couplings in the low-energy effective
theory.

Finally, we would  like to understand how partial breaking
works in the presence of gravity.  Gravity couples to the
true stress-energy tensor, so it distinguishes between the
right-hand sides of the commutators (6).  Some early work
on this question was done by Cecotti, Girardello and
Porrati (1986) and by Zinov'ev (1987).  These groups
considered nonminimal cases and found that their gravitational
couplings utilize the second loophole.  One would like to
reconcile their results with those above.

\section{Coset Construction}

In this talk we will take a bottom-up approach to the subject
of partial supersymmetry breaking.  We will use nonlinear
realizations to describe the effective $N=1$ theory which holds
below the scale of the second supersymmetry breaking.  We will
use the formalism of Coleman, Wess and Zumino (1969), as
extended by Volkov (1973), to construct theories where
the $N=1$ supersymmetry is manifest, and the second supersymmetry
is nonlinearly realized.

The approach of Coleman, Wess, Zumino and Volkov is based on a
coset decomposition of a symmetry group, $G$.  We start with
a group, $G$, of internal and spacetime symmetries, and
partition the generators of $G$ into three classes:
\begin{itemize}
\item
$\Gamma_A$, the generators of unbroken spacetime translations;
\item
$\Gamma_a$, the generators of spontaneously broken internal and
spacetime symmetries; and
\item
$\Gamma_i$, the generators of unbroken spacetime rotations
and unbroken internal symmetries.
\end{itemize}
The generators $\Gamma_i$ close into the stability group, $H$.

Given $G$ and $H$, we define the coset $G/H$ in terms of
an equivalence relation on the elements $\Omega \in G$,
$\Omega \sim \Omega\, h$, with $h \in H$.  The coset can be
thought of as a section of a fiber bundle with total space,
$G$, and fiber, $H$.

This equivalence relation suggests that we parametrize
the coset as follows,
\begin{equation}
\Omega\ =\ \exp \I X^A \Gamma_A\ \exp \I\xi^a(X) \Gamma_a\ .
\end{equation}
Physically, the $X^A$ play the role of generalized spacetime
coordinates, while the $\xi^a(X)$ are generalized Goldstone
fields, defined on the generalized coordinates and valued in
the set of broken generators $\Gamma_a$.  There is one
generalized coordinate for every unbroken spacetime translation,
and one generalized Goldstone field for every spontaneously
broken generator.

We define the action of the group $G$ on the coset $G/H$ by
left multiplication, $\Omega \rightarrow g\, \Omega = \Omega^\prime
\, h$, with $g \in G$.  In this expression,
\begin{equation}
\label{define primes}
\Omega^\prime\ =\ \exp \I X^{\prime A} \Gamma_A
\ \exp \I \xi^{\prime a}(X^\prime) \Gamma_a
\end{equation}
and $h = \exp \I \alpha^i (g, X, \xi) \Gamma_i.$
The group multiplication induces nonlinear transformations
on the coordinates $X^A$ and the Goldstone fields $\xi^a$:
\be
X^A\ \rightarrow\ X^{\prime A}\ , \quad
\xi^a(X)\ \rightarrow\ \xi^{\prime a}(X^\prime)\ .
\ee
These transformations realize the full symmetry group, $G$.  Note
that the field $\xi^a$ transforms by a shift under the transformation
generated by $\Gamma_a$.  This confirms that $\xi^a$ is indeed the
Goldstone field corresponding to the broken generator $\Gamma_a$.

An arbitrary $G$ transformation induces a compensating $H$
transformation which is required to restore the section.  This
transformation can be used to lift any representation, $R$, of
$H$, to a nonlinear realization of the full group, $G$, as
follows,
\begin{equation}
\chi(X)\ \rightarrow\ \chi^\prime(X^\prime)\ =\ D(h)
\chi(X)\ .
\end{equation}
Here $D(h) = \exp(\I \alpha^i T_i)$, where $\alpha^i$ was defined
below (\ref{define primes}), and the $T_i$ are generators of $H$ in
the representation $R$.

To proceed further, it is helpful to have a vielbein, connection
and covariant derivative, built from the Goldstone fields in the
following way.  One first computes the Maurer-Cartan
form, $\Omega^{-1} \d\Omega$, where d is the exterior derivative.
One then expands $\Omega^{-1} \d\Omega$ in terms of the Lie algebra
of $G$,
\begin{equation}
\Omega^{-1} \d \Omega\ =\ \I (\omega^A \Gamma_A +
\omega^a \Gamma_a + \omega^i\Gamma_i)\ ,
\label{omegas}
\end{equation}
where $\omega^A, \omega^a$ and $\omega^i$ are one-forms on the
manifold parametrized by the coordinates $X^A$.

The Maurer-Cartan form transforms as follows under a rigid
$G$ transformation,
\begin{equation}
\Omega^{-1} d \Omega\ \rightarrow\ h (\Omega^{-1} \d \Omega) h^{-1}
 -\ \d h\,h^{-1}\ .
\end{equation}
{}From this we see that the fields $\omega^A$ and $\omega^a$
transform covariantly under $G$, while $\omega^i$ transforms by
a shift.  These transformations help us identify
\begin{equation}
\omega^A\ =\ \d X^M \, E_M{}^A
\end{equation}
as the covariant vielbein,
\begin{equation}
\omega^a\ =\ \d X^M \, E_M{}^A \cD_A \xi^a
\end{equation}
as the covariant derivative of the Goldstone field $\xi^a$,
and
\begin{equation}
\omega^i\ =\ \d X^M \, \omega_M^i
\label{h connection}
\end{equation}
as the connection associated with the stability group, $H$.
With these building blocks, it is easy to construct theories
invariant under the full group $G$.

The coset construction is very general and very powerful.
For the case of internal symmetries, it allows one to prove
that any $H$-invariant action can be lifted to be $G$-invariant
with the help of the Goldstone bosons.  For $N=1$ supersymmetry,
it can be used to show that any Lorentz-invariant action can
be made supersymmetric with the help of the Goldstone
fermion.

\section{Nonlinear Supersymmetry}

In this section we will show that any $N=1$ supersymmetric
theory can be made $N=2$ supersymmetric with the help of an
$N=1$ Goldstone superfield.  We will find that the Goldstone
superfield can contain either an $N=1$ chiral or vector multiplet
(Bagger and Galperin, 1994, 1997a).
[The case where the Goldstone superfield is an $N=1$ tensor
multiplet can be obtained from the chiral case by a superspace
duality transformation (Bagger and Galperin, 1997b).]

It is important to emphasize that the coset construction --
while very useful and very general -- does not tell us
anything about the underlying theory in which both
supersymmetries are linearly realized.  Indeed, such
a theory might not even exist.  Therefore we shall resolutely
insist that we are working in the context of an effective field
theory, and leave to others the task of finding the more
fundamental theory above the supersymmetry-breaking scale.

In what follows we shall first take a minimal approach, and choose
the group $G$ to be the supergroup whose algebra is (\ref{susyalg}).
We will take the subgroup $H$ to be the supergroup generated
by $P_a$, $Q_\alpha$ and $\bQ_{\dalpha}$.  We parametrize the coset
element $\Omega$ as follows,
\bea
\label{real_parametrization}
\Omega &\ =\ & \exp \I(x^aP_a +\theta^\alpha Q_\alpha +\bar\theta_{\dot\alpha}
\bar Q^{\dot\alpha})\nn\\
&&\quad\times
\exp \I(\Psi^\alpha S_\alpha +\bar\Psi_{\dot\alpha}\bar S^{\dot\alpha})\ .
\eea
Here $x,\ \theta$ and $\bar\theta$ are the coordinates of $N=1$
superspace, while $\Psi^\alpha$ and its conjugate $\bar\Psi_{\dot\alpha}$
are Goldstone $N=1$ superfields of (geometrical) dimension $-1/2$.
These spinor superfields contain far too many component fields, so
we need to find a set of consistent, covariant constraints to reduce
the number of fields.

The correct constraints are most easily expressed in term of the
$N=2$ covariant derivatives of the Goldstone superfield.  The covariant
derivatives can be found following the techniques of the previous
section; they can be explicitly written as follows,
\begin{eqnarray}
\label{derivatives}
{\cal D}_\alpha &\ =\ & D_\alpha
 - \I(D_\alpha\Psi\sigma^a\bar\Psi +
 D_\alpha\bar\Psi\bar\sigma^a\Psi)\omega^{-1}_a{}^m\partial_m \nn \\
\bar{\cal D}_{\dot\alpha} &=& \bar D_{\dot\alpha} -
\I(\bar D_{\dot\alpha}\Psi\sigma^a\bar\Psi +
\bar D_{\dot\alpha}\bar\Psi\bar\sigma^a\Psi)\omega^{-1}_a{}^m
\partial_m\nn\\
{\cal D}_a &=& \omega^{-1}_a{}^m\partial_m\ ,
\end{eqnarray}
where $\omega_m{}^a \equiv \delta_m^a + \I(\partial_m\Psi\sigma^a\bar\Psi
+\partial_m\bar\Psi\bar\sigma^a\Psi)$ and $D_\alpha,\ \bar D_{\dot\alpha}$
are ordinary flat $N=1$ superspace spinor derivatives.  The
covariant derivatives obey the following commutation relations,
\begin{eqnarray}
\label{cov_algebra}
\{ {\cal D}_\alpha, {\cal D}_\beta \} &\ = \ &
- 2\I({\cal D}_\alpha\Psi^\gamma
{\cal D}_\beta\bar\Psi^{\dot\gamma} +
(\alpha \leftrightarrow \beta))
{\cal D}_{\gamma\dot\gamma} \nn\\
\left[ {\cal D}_\alpha, {\cal D}_a \right] & = &
-2\I ({\cal D}_\alpha \Psi^\gamma
{\cal D}_a \bar\Psi^{\dot\gamma} +
(\alpha\leftrightarrow a))
{\cal D}_{\gamma\dot\gamma} \nn \\
\{ {\cal D}_\alpha, \bar{\cal D}_{\dot\beta} \} & = &
\ 2\I\sigma^a_{\alpha\dot\beta}{\cal D}_a
-2\I({\cal D}_\alpha\Psi^\gamma
\bar{\cal D}_{\dot\beta}\bar\Psi^{\dot\gamma} \nn\\
&& \quad + (\alpha \leftrightarrow \dot\beta ))
{\cal D}_{\gamma\dot\gamma}\ ,
\end{eqnarray}
where ${\cal D}_{\alpha \dalpha} \equiv \sigma^a_{\alpha \dalpha}
{\cal D}_a$.

One set of constraints is simply (Bagger and Galperin, 1994)
\bea
\cbD\cbD\,\Psi_\alpha &\ =\ & {\cal O}(\Psi^3) \nn\\
\cD_\alpha \Psi_\beta + \cD_\beta \Psi_\alpha
 &=& {\cal O}(\Psi^3)\ .
\label{chiral constraints}
\eea
The right-hand side of this equation must be adjusted for
consistency with (\ref{cov_algebra}).  Remarkably, this can
be done using the dimensionless invariants
$\bar{\cal D}_{\dot\alpha} \Psi_\alpha$ and
${\cal D}_\alpha\Psi_\beta$
(together with their complex conjugates).  It turns
out that there is a unique, consistent solution order-by-order
in powers of the Goldstone field.

The solution to the constraints (\ref{chiral constraints}) is
easy to find in perturbation theory.  To lowest order, it is
just the chiral multiplet $\Phi$,
\bea
\Psi_\alpha &\ =\ & D_\alpha \Phi + {\cal O}(\Psi^3)  \nn\\
\bD_{\dalpha} \Phi &=& {\cal O}(\Psi^3)\ .
\eea
In this expression, $D_\alpha$ is the ordinary $N=1$ superspace
spinor derivative.

A second set of constraints is (Bagger and Galperin, 1997a)
\bea
\cbD_{\dalpha} \Psi_\alpha &\ =\ & {\cal O}(\Psi^3) \nn\\
\cD^\alpha \Psi_\alpha + \cbD_{\dbeta} \bar\Psi^{\dbeta}
 &=& {\cal O}(\Psi^3)\ .
\label{vector constraints}
\eea
As above, the right-hand side must be adjusted for
consistency with the algebra of covariant derivatives.
Again, there is a unique, consistent solution.  To
lowest order in perturbation theory, it is
\bea
\Psi_\alpha &\ =\ & W_\alpha + {\cal O}(\Psi^3)  \nn\\
W_\alpha &=& - {1\over4}\bD\bD D_\alpha V + {\cal O}(\Psi^3)\ ,
\eea
where $V$ is a real $N=1$ vector superfield.  We see
that the chiral and vector Goldstone multiplet can each
be obtained to lowest order in perturbation theory.
In fact, the consistency of the multiplets survives
to all orders in perturbation theory.

The Goldstone action can be constructed order-by-order in the
Goldstone fields.  For the chiral case, it is simply
(Bagger and Galperin, 1994)
\be
\label{goldaction1}
S\ =\ v^4 \int d^4x d^2\theta d^2\bar\theta \,E\,
[\Phi^+ \Phi + {\cal O}(\Phi^4)]\ .
\ee
In this expression, $E =$Ber($E_{M}{}^{A}$) is the superdeterminant
of the vielbein, and $v$ is the constant of dimension one which
corresponds to the scale of the supersymmetry breaking.
The action (\ref{goldaction1}) is invariant under the full $N=2$
supersymmetry.

For the vector multiplet, the Goldstone action is just
(Bagger and Galperin, 1997a)
\bea
\label{goldaction2}
S &\ =\ & {v^4\over 4} \int d^4x d^2 \theta \,{\cal E}\, W^2 \plushc \nn\\
&& \quad + \int d^4x d^4\theta \,E\, {\cal O}(W^4)\ .
\eea
This action is invariant under $N=2$ supersymmetry.  It is also
gauge-invariant.  Curiously enough, the gauge field contribution to
the Goldstone action coincides with the expansion of the Born-Infeld
action.

Having constructed the $N=2$ Goldstone action, we are now ready to
add $N=2$ covariant matter.  The basic ingredients are $N=2$ nonlinear
generalizations of $N=1$ chiral and vector superfields.  The generalized
chiral superfields are defined by the constraint $\cbD_{\dalpha}\chi =
0$.  This constraint is consistent for either type of Goldstone
multiplet.

The matter action is easy to write down for either Goldstone
multiplet.  The kinetic term is
\be
S\ =\ \int d^4x d^4\theta  \,E \, K(\chi^+,\chi)
\ee
while the superpotential term is
\be
S\ =\ \int d^4x d^2\theta \,{\cal E}\,P(\chi)\ .
\ee
As before, $E$ and ${\cal E}$ are superdeterminants of the
supervielbein $E_M{}^A$.  They can be adjusted to preserve the
condition
\be
\int d^4x d^4\theta \,E\, F(\chi)\ =\ 0\ .
\ee
This allows the matter action to be K\"ahler invariant, so the
matter couplings are described in terms of K\"ahler manifolds,
just as for $N=1$.

It is not hard to generalize these results to include vector
superfields.  The general conclusion is that any $N=1$ invariant
theory can be lifted to be $N=2$ supersymmetric with the help of a
Goldstone superfield.  Furthermore, the Goldstone superfield can
be either an $N=1$ chiral or vector multiplet.

Now that we have two explicit realizations of partial supersymmetry
breaking, we can ask how they avoid the no-go argument
discussed above.  In each case, the nonlinear theory exploits the
loophole of Hughes, Liu and Polchinski (1986).  For example, in the
vector case the second supercurrent goes like $J^m_\alpha \sim v^4
\sigma^m_{\alpha\dalpha} \blambda^{\dalpha}$, so its commutator
with the second supercharge reproduces the algebra (\ref{current algebra}).

\section{Geometry}

The fact that the constraints need to be adjusted order-by-order
in $\Psi_\alpha$ hints that a deeper structure underlies
partial supersymmetry breaking.  The $N=2$ supersymmetry does not
provide enough symmetry to uniquely fix the covariant derivatives
and the associated constraints.  This intuition is borne
out for the case of the chiral multiplet, where a much deeper set
of symmetries acts on the Goldstone multiplet (Bagger and Galperin,
1994).

To see this, let us first extend the $N=2$ algebra by a complex
central charge, $Z$:
\bea
\{Q_\alpha, \bar Q_{\dot\alpha}\}\ =
\ 2\sigma^a_{\alpha\dot\alpha}P_a &\qquad&
\{S_\alpha, \bar S_{\dot\alpha}\}\ =
\ 2\sigma^a_{\alpha\dot\alpha}P_a \nn\\
\{Q_\alpha, S_\beta\}\ = \ 2\epsilon_{\alpha\beta}Z\  \ &&
\{\bar Q_{\dot\alpha}, \bar S_{\dot\beta}\}\ =
\ 2\epsilon_{\dot\alpha\dot\beta}\bar Z\ .
\label{susy alg 2}
\eea
We then consider a coset where the group $G$ contains
not only $N=2$ supersymmetry, but also its maximal
automorphism group, $SO(5,1) \times SU(2)$, where the $SU(2)$
acts on the two supersymetry generators, and $SO(5,1)$ is the
$D=6$ Lorentz group.  (Under $SO(5,1)$, the generators $P_a$ and
$Z$ form a $D=6$ vector, while the two supercharges form a single
$D=6$ Majorana-Weyl spinor).  Let us take $H$ to be $SO(3,1)\times
SO(2) \times U(1)$, where $SO(3,1)\times SO(2) \subset SO(5,1)$,
$U(1)\subset SU(2)$, and $SO(3,1)$ is the $D=4$ Lorentz group.

Our parametrization of the coset $G/H$ involves the $N=1$
superspace coordinates, as well as different Goldstone
superfields for each of the broken symmetries,
\bea
\Omega &\ =\ & \exp \I(x^aP_a+\theta^\alpha Q_\alpha
+\bar\theta_{\dot\alpha}\bar Q^{\dot\alpha}) \nn\\
&&\times
\exp \I(\Phi Z +\bar\Phi\bar Z +
\Psi^\alpha S_\alpha+\bar\Psi_{\dot\alpha}\bar S^{\dot\alpha}) \nn\\
&&\times \exp \I(\Lambda^aK_a+ \bar\Lambda^a\bar K_a +\Xi T
+\bar\Xi\bar T)\ .
\eea
Here $\Lambda^a,\, \bar\Lambda^a$ are the Goldstone superfields
associated with the generators $K_a,\, \bar K_a$ of $SO(5,1)/
SO(3,1)\times SO(2)$.  Similarly, $\Xi,\, \bar\Xi$ are the
Goldstone superfields for the broken generators
$T,\, \bar T$ of $SU(2)/U(1)$.

As before, the $N=1$ Goldstone superfields contain far more components
than the minimal Goldstone multiplet.  This motivates us to impose
the following consistent set of constraints:
\bea
{\bar{\cal D}}_{\dot\alpha}\Phi\ =\ 0\ ,&&
\qquad {\cal D}_{\alpha}\Phi\ =\ 0\ ,
\qquad {\cal D}_{a}\Phi\ =\ 0 \nn \\
{\cal D}_{\alpha}\Psi^\beta\ =\ 0\ ,&&
\qquad {\bar{\cal D}}_{\dot\alpha}\Psi^\beta\ =\ 0\ .  \label{constr2}
\eea
These constraints allow us to express the Goldstone superfields
$\Psi^\alpha, \Lambda^a$ and $\bar\Xi$ in terms of a single superfield
$\Phi$.  [This way of eliminating Goldstones was called the ``inverse
Higgs effect" by Ivanov and Ogievetsky (1975).]  To lowest order,
we find $\Psi^\alpha = -{\I\over 2}D^\alpha\Phi$,
$\Lambda_a = -\partial_a\Phi$, and
$\bar\Xi = {1\over 4}D^2\Phi$.
The constraint ${\bar{\cal D}}_{\dot\alpha}\Phi=0$ reduces
$\Phi$ to an $N=1$ chiral superfield.

The remarkable fact about this construction is that it reveals
a geometrical role for each component of the chiral Goldstone
multiplet.  The scalar field, $A$, is the complex Goldstone boson
associated with the spontaneously broken central charge symmetry.
Its derivative, $\partial_m A$, is the Goldstone boson associated
with $SO(5,1)/SO(3,1)\times SO(2)$.  The $F$-component of $\Phi$
is the complex Goldstone boson associated with the $SU(2)/U(1)$.
Finally, the spinor is the Goldstone fermion that arises
from the partially broken supersymmetry.

The action (\ref{goldaction1}) turns out to be invariant under
$SO(5,1)$, but it explicitly breaks $SU(2)$ down to $U(1)$.
Furthermore, any $R$-invariant $N=1$ matter action can
be lifted to be $SO(5,1)$ invariant.
These facts hint that the Goldstone action might be related
to the six-dimensional membrane of Hughes, Liu and Polchinski
(1986).
Indeed, it is not hard to show that the chiral Goldstone action
is precisely the gauge-fixed membrane action.

The geometry that underlies the vector case is presently under
study.  The Born-Infeld form of the gauge action suggests that
it might be related to some sort of D-brane.  The fact that
there are no ``transverse" scalars hints that the action
might be that of a space-filling D3-brane.  In any case,
one would like to find the Goldstone-type symmetries associated
with the gauge field strength and the auxiliary field of the
Goldstone multiplet.

In fact, the $D$-component of the Goldstone multiplet
can be interpreted as the Goldstone boson associated with
the following $U(1)$ subgroup of the $SU(2)$ automorphism
symmetry:
$\delta\theta^\alpha = \I\eta\Psi^\alpha$,
$\delta\Psi^\alpha = \I\eta\theta^\alpha$.
Under such a transformation, the $D$-component is shifted by
the constant parameter $\eta$.

If we were to extend $G$ in $G/H$ by this $U(1)$, we would
eliminate the dimensionless invariant ${\cal D}^\alpha\Psi_\alpha$
in favor of the corresponding Goldstone superfield.  Even then,
there would still be a dimensionless invariant associated with
the gauge field strength, ${\cal D}_{(\alpha}\Psi_{\beta)}$.
This suggests that there is an extension of $N=2$ supersymmetry
which associates a Goldstone-like symmetry with this field
strength.

Moreover, gauge fields themselves can be interpreted as
Goldstone fields associated with infinite-dimensional symmetry
groups (Ivanov and Ogievetsky, 1976).
This leads us to wonder whether the full symmetry of
the new multiplet is some infinite-dimensional extension of
$N=2$ supersymmetry.

\section{Supergravity}

We have just seen that there are two independent Goldstone
realizations of partial supersymmetry breaking in four
dimensions.  (A third is related by duality.)  Both give
rise to the current algebra (\ref{current algebra}).  Because
the spontaneous breaking relies on the curious shift
in the ``second" stress-energy tensor, one would like to
see what happens when the Goldstone multiplets are
coupled to supergravity.

In this section, we will work backwards, and start by
constructing two Lagrangians and two sets of supersymmetry
transformations for the massive $N=1$ spin-3/2 multiplet.
We will then ``unHiggs" the theories by adding appropriate
Goldstone fields and coupling gravity.  In this way we
will find the supergravities associated with each of the
Goldstone multiplets.  (The work in this section was
done in collaboration with Richard Altendorfer and Samuel
Osofsky.)

We will see that the second Lagrangian corresponds to an
alternative representation for the $N=1$ massive spin-3/2
multiplet, one which was originally found by Ogievetsky and
Sokatchev (1977).  When coupled to gravity, this representation
gives rise to a new $N=2$ supergravity with a modified
$N=2$ supersymmetry algebra.

\subsection{The Massive $N=1$ Spin-3/2 Multiplet}

The starting point for the supergravity coupling is the massive
$N=1$ spin-3/2 multiplet.  This multiplet contains six bosonic
and six fermionic (on-shell) degrees of freedom, arranged in states of the
following spins,
\be
\pmatrix{
{3\over2} \crr
1\ \ \ 1 \crr
{1\over2}}\ .
\ee
The traditional representation of this multiplet contains
the following fields (Ferrara and van Nieuwenhuizen, 1983):
one spin-3/2 fermion, one spin-1/2
fermion, and two spin-one vectors, each of mass $m$.  The
Ogievetsky-Sokatchev representation has the same fermions, but
just one vector plus one antisymmetric tensor.  As
we shall see, each representation has a role to play in the
theory of partial supersymmetry breaking.

The traditional representation is described by the following
Lagrangian (Ferrara and van Nieuwenhuizen, 1983):
\bea
\cL &\ = \ & \epsilon^{m n \rho \sigma} \overline \psi_{m}
  \overline \sigma_n \partial_\rho \psi_\sigma 
 - \I \overline \zeta \overline \sigma^m \partial_m \zeta
 - {1 \over 4} \cA_{m n} \bar\cA^{m n} \nonumber \\
& &-\ {1\over 2}m^2\, \cA_m \bar\cA^m  
\ +\ {1\over 2}m\,\zeta\zeta  \ +\ {1\over 2}m\,\bar\zeta\bar\zeta \nonumber \\[1mm]
& & -\ m\,\psi_m \sigma^{m n} \psi_n -\ m\,\bar\psi_m \bar\sigma^{m n} \bar\psi_n\ .
\eea
Here $\psi_m$ is a spin-3/2 Rarita-Schwinger field, $\zeta$ a spin-1/2
fermion, and $\cA_m = A_m + \I B_m$ a complex spin-one vector.  This
Lagrangian is invariant under the following $N=1$ supersymmetry
transformations,
\bea
\delta_\eta \cA_m &\ =\ & 2\psi_m\eta - \I{2\over\sqrt{3}}
\bar\zeta\bar\sigma_m\eta 
-{2\over \sqrt{3}m}\partial_m(\zeta\eta) \nonumber \\
\delta_\eta \zeta &=& {1\over\sqrt{3}}\bar\cA_{mn}\sigma^{mn}
\eta -\I{m\over\sqrt{3}}
\sigma^m\bar\eta \cA_m \nonumber \\
\delta_\eta \psi_m &=& {1\over 3m}\partial_m(\bar\cA_{rs}
\sigma^{rs} \eta + 2\I m 
\sigma^n\bar\eta \cA_n)  
- {\I\over 2}(H_{+mn}\sigma^n + {1\over 3}H_{-mn}\sigma^n)
\bar\eta \nonumber
\\ & & -\ {2\over 3}m({\sigma_m}^n \bar\cA_n \eta + \bar
\cA_m\eta)\ ,
\eea
where $H_{\pm mn}=\cA_{mn}\pm {\I\over 2}\epsilon_{mnrs}\cA^{rs}$.

The alternative Ogievetsky-Sokatchev representation has the following
Lagrangian,
\bea
\cL &\ =\ & \epsilon^{pqrs} \bar \psi_{p}
  \bar \sigma_q \partial_r \psi_s
 - \I \bar \zeta \bar \sigma^m \partial_m \zeta
 - {1 \over 4} A_{mn} A^{mn} 
 + {1\over 2}v^m v_m \nonumber \\
& & -\ {1\over 2}m^2 A_m A^m  - {1\over4}m^2
B_{mn}B^{mn} \ +\ {1\over 2}m\,\zeta\zeta  \ +\ {1\over
2}m\,\bar\zeta\bar\zeta\nonumber\\[1mm]
& & -\ m\,\psi_m \sigma^{mn} \psi_n \ -\ m\,\bar\psi_m \bar\sigma^{mn}
\bar\psi_n\ ,
\eea
where $A_{mn}$ is the field strength associated with the real vector
field $A_m$, and $v_m = {1\over 2}\epsilon_{mnrs} \partial^n B^{rs}$ is
the field strength for the antisymmetric tensor
$B_{mn}$.  This Lagrangian is invariant under the following
$N=1$ supersymmetry transformations,
\bea
\delta_\eta A_m &\ =\ & (\psi_m\eta + \bar\psi_m\bar\eta) + 
{\I\over\sqrt{3}}
(\bar\eta\bar\sigma_m\zeta - \bar\zeta\bar\sigma_m\eta)
-{1\over \sqrt{3}m}\partial_m(\zeta\eta +  \bar\zeta\bar\eta) 
\nonumber \\
\delta_\eta B_{mn} &=& {2\over\sqrt{3}}\left( \eta\sigma_{mn}\zeta 
+ {\I\over 2m}\partial_{[m}\bar\zeta\bar\sigma_{n]}\eta \right) 
\ +\ \I\eta\sigma_{[ m}\bar\psi_{n ]} + {1\over
m} \eta\psi_{mn}
\plushc \nonumber\\
\delta_\eta \zeta &=& {1\over\sqrt{3}} A_{mn}\sigma^{mn}\eta - 
{\I m\over\sqrt{3}}
 \sigma^m\bar\eta A_m - {1\over\sqrt{3}}m\sigma_{mn}\eta B^{mn} - 
{1\over\sqrt{3}} v_m\sigma^m\bar\eta \nonumber \\
\delta_\eta \psi_m &=& {1\over 3m}\partial_m \left( A_{rs}
\sigma^{rs}\eta + 2\I m
\sigma^n\bar\eta A_n  \right) 
 -{\I\over 2} (H^A_{+mn}\sigma^n + {1\over 3}H^A_{-mn}\sigma^n)
\bar\eta \nonumber \\
& & -\ {2\over 3}m ({\sigma_m}^n A_n \eta +
A_m\eta)\ +\ {1\over 3m}\partial_m \left( 2
v_n\sigma^n\bar\eta  - m 
 \sigma^{rs}\eta B_{rs} \right) \nonumber\\
&&    -\ {2\I\over 3} (v_m + \sigma_{mn}v^n)\eta 
 - {\I m\over 3} (B_{mn}\sigma^n\bar\eta + \I
\epsilon_{mnrs}B^{n r}\sigma^s\bar\eta) \ ,
\eea
where the square brackets denote antisymmetrization, without a
factor of 1/2.

These Lagrangians describe the free dynamics of massive spin-3/2
and 1/2 fermions, together with their supersymmetric partners,
massive spin-one vector and tensor fields.  They can be thought
of as ``unitary gauge" representations of theories with additional
symmetries:  a second supersymmetry for the massive spin-3/2 fermion,
and additional gauge symmetries associated with the massive gauge
fields.

\subsection{The Supergravity Coupling}

To study partial breaking, we need to ``unHiggs" these Lagrangians
by including appropriate gauge and Goldstone fields.  In each case
we need to add a Goldstone multiplet and gauge the full $N=2$
supersymmetry.  The supersymmetric partners of the Goldstone
fermion will turn out to be the Goldstone bosons that restore
the gauge symmetries associated with the massive bosonic fields.
At the end of the day, we will find two theories with $N=2$
supersymmetry nonlinearly realized, but $N=1$ represented linearly
on the fields. The resulting effective field theories describe
the physics of partial supersymmetry breaking, well below the
scale where the second supersymmetry is broken.

The trick to this construction is to add the right fields.  Because
$N=1$ supersymmetry is not broken, the Goldstone fermion must belong
to an $N=1$ supersymmetry multiplet.  For the two cases of interest,
we shall see that the Goldstone fermion must belong to the chiral or
the vector multiplet, discussed above.

Let us first consider the chiral case.  Under the first supersymmetry,
a complex boson $\phi$ transforms into a Weyl fermion $\chi$,
\be
\delta_{\eta^1}\phi \ =\ \sqrt 2\,\eta^1\chi\ .
\ee
If $\chi$ is the Goldstone fermion, it shifts under the second
supersymmetry,
\be
\delta_{\eta^2}\chi\ =\ \sqrt 2\,v^2\,\eta^2\ +\ \ldots\ ,
\ee
where $v$ is the scale of the second supersymmetry breaking.  Therefore
the closure of the two supersymmetries on $\phi$ gives
\be
[\,\delta_{\eta^2},\,\delta_{\eta^1}\,]\,\phi\ =\ 2\,v^2\,\eta^1\eta^2\ +
\ \ldots
\ee
The complex scalar $\phi$ undergoes a constant shift.  This is in accord
with our previous result:  The field $\phi$ is itself a Goldstone boson,
corresponding to a complex central charge.  It expects to be eaten by a
complex vector field, which suggests that the chiral Goldstone multiplet
should be associated with the traditional representation for the massive
spin-3/2 multiplet.

As shown in Figure 1(a), the degree of freedom counting works out just
right.  We start with the $N=1$ chiral Goldstone multiplet and add an
$N=1$ vector multiplet.  We then add the gauge fields of $N=2$ supergravity.
As we will see, the full set of fields can be used to construct a
Lagrangian which is invariant under $N=2$ supersymmetry.  The final
results look complicated, but they are actually very simple:  In unitary
gauge, the two vectors eat the two scalars, while the Rarita-Schwinger
field eats one linear combination of the spin-1/2 fermions.  This leaves
the massive $N=1$ multiplet coupled to $N=1$ supergravity.

\begin{figure}[t]
\hspace*{0.1truein}
\psfig{figure=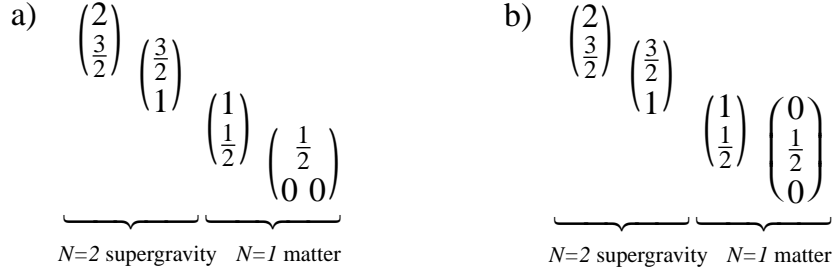,height=1.4in}
\caption{The unHiggsed versions of the (a) traditional and (b)
alternative representations of the $N=1$ massive spin-3/2 multiplet.
The traditional representation contains the degrees of freedom
associated with an $N=1$ chiral multiplet.  The alternative
representation exchanges the chiral multiplet for its dual,
an $N=1$ tensor multiplet.}
\end{figure}

With that said, we now present the Lagrangian (Altendorfer, Bagger,
Osofsky, 1998):
\bea
&& e^{-1}\cL \ =\ \nn\\
&& -\ {1 \over 2 \kappa^2} {\cal R}
 + \epsilon^{m n r s} \overline \psi_{m i}
  \overline \sigma_n D_r \psi^i_s 
 - \I \overline \chi \  \overline \sigma^m D_m \chi 
 - \I \overline \lambda \overline \sigma^m D_m \lambda
-  \cD^m \phi \overline{\cD_m \phi} \nn\\
&& -\ {1 \over 4} \cA_{m n} \overline \cA^{m n} 
 - \ \Bigl( {1 \over \sqrt{2}} m  \psi^2_m \sigma^m
   \overline \lambda 
 + \I m\psi^2_m \sigma^m
   \overline \chi 
+ \sqrt{2}  \I m \lambda \chi 
 +\ {1 \over 2} m \chi \chi \nn\\
&& +\ \Red m \Black\, \psi^2_m \sigma^{m n} \psi^2_n 
+\ {\kappa \over 4} \epsilon_{i j}
\psi^i_m \psi_{n}^{j}
    \overline H_+^{m n} 
 + {\kappa \over \sqrt{2}}  \chi \sigma^m
    \overline \sigma^n \psi^1_m \overline{\cD_n \phi} 
 \nonumber \\
& & + \  {\kappa \over 2 \sqrt{2}}  \overline
\lambda
    \overline \sigma_m \psi^1_n \overline H_-^{m n}
 +  {\kappa \over \sqrt{2}}
    \epsilon^{m n r s} 
    \overline \psi_{m 2} \overline \sigma_n \psi^1_r
    \overline{\cD_s \phi}
    \plushc \Bigr) \ ,
\eea
where $\kappa$ denotes Newton's constant, $m = \kappa v^2$, and
\bea
\cA_m&\ =\ &A_m + \I B_m \nonumber \\
\cA_{mn}&=&\partial_m \cA_m - \partial_n \cA_m \nonumber\\
H_{\pm mn}&=&\cA_{mn}\pm {\I\over
2}\epsilon_{mnrs}\cA^{rs}\ .
\eea
The supercovariant derivatives are as follows,
\bea
\hat\cD_m\phi&\ =\ &\partial_m\phi - {\kappa\over\sqrt{2}}\psi^1_m\chi - 
{1\over\sqrt{2}} \kappa v^2  \cA_m\nonumber \\
\hat\cA_{mn}&=&\cA_{mn} + \kappa\psi^2_{ [m}\psi^1_{n] } - 
{\kappa\over\sqrt{2}}\bar\lambda\bar\sigma_{ [n}\psi^1_{m ]} \ .
\eea
This Lagrangian is invariant under two independent abelian gauge
symmetries, as well as the following supersymmetry transformations,
\bea
\delta e^a_m &\ =\ &\I \kappa (\eta^i \sigma^a \overline \psi_{m i} + 
   \bar\eta_i \bar\sigma^a \psi_{m}^i) \nonumber \\
\delta \psi^i_{m} & = & {2 \over \kappa} D_m \eta^i
     \nonumber \\
& &   +\ \left( -{\I \over 2} \hat H_{+m n}
\sigma^n \overline \eta_1
 + \sqrt{2} \overline{\cD_m \phi} \eta^1 
 - \kappa\psi^1_m(\bar\chi\bar\eta_1) 
 + \I v^2 \sigma_m \overline \eta_2 \right){\delta_2}^i \nonumber \\
\delta \cA_m &=& 2 \epsilon_{i j} \psi_{m}^{i} \eta^j 
+ \sqrt{2}  \overline \lambda \overline \sigma_m \eta^1 \nonumber \\
\delta \lambda  &=&  {\I \over \sqrt{2}} \overline{\hat\cA}_{mn} 
\sigma^{mn} \eta^1
\Red - \I \sqrt{2} v^2 \eta^2 \Black \nonumber \\
\delta \chi  &=&  \I \sqrt{2} \sigma^m {\hat\cD_m \phi}\overline \eta_1
\Red + 2 v^2 \eta^2 \Black \nonumber \\
\delta \phi &=& \sqrt{2} \chi \eta^1 \ ,
\eea
for $i=1,2$.  This result holds to leading order, that is, up
to and including terms in the transformations that are linear in
the fields.  Note that this representation is irreducible in the sense
that there are no subsets of fields that transform only into themselves
under the supersymmetry transformations.  (Because of this, the
multiplet structure outlined in Fig.~1 is slightly misleading.)

Let us now consider the case of the vector Goldstone multiplet.  Under
the first supersymmetry, the real vector $B_m$ of a vector multiplet 
transforms into a Weyl fermion $\lambda$,
\be
\delta_{\eta^1} B_m\ =\ \sqrt 2\I\,(\lambda\sigma_m\bar\eta^1 -
\eta^1\sigma_m\bar\lambda)\ .
\ee
If $\lambda$ is the Goldstone fermion, it shifts under the second
supersymmetry.  Therefore the closure of the two supersymmetries
on $B_m$ gives
\be
[\,\delta_{\eta^2},\,\delta_{\eta^1}\,]\,B_m\ =\ 2\I v^2\,
(\eta^2\sigma_m\bar\eta^1 -
\eta^1\sigma_m\bar\eta^2)\ +
\ \ldots
\ee
From this we see that the real vector $B_m$ is a Goldstone
boson.  It expects to be eaten by an antisymmetric tensor field.
This suggests that the vector Goldstone multiplet should
be associated with the alternative representation for the
massive spin-3/2 multiplet.

The degree of freedom counting is shown in Figure 1(b).  As before,
we include the $N=2$ supergravity multiplet.  This time, however,
the matter fields include the $N=1$ vector Goldstone multiplet,
together with an $N=1$ tensor multiplet.  In unitary gauge, one
vector eats one scalar, while the antisymmetric tensor eats the
other vector.  [The massless antisymmetric tensor field contains
one degree of freedom.  It was introduced by Ogievetsky and
Polubarinov (1966), who called it the ``notoph," or inverse
photon.]  These are the minimal set of fields that arise
when coupling the Ogievetsky-Sokatchev spin-3/2 multiplet to
$N=2$ supergravity.

The Lagrangian for this system can be worked out following the
same procedure described above.  One finds (Altendorfer, Bagger,
Osofsky, 1998):
\bea
& &e^{-1}\cL\ =\ \nonumber \\
&  &  -\ {1 \over 2 \kappa^2} {\cal R}
 + \epsilon^{pqrs} \bar \psi_{p i}
  \bar \sigma_q D_r \psi^i_s 
 - \I \bar \chi \bar \sigma^m D_m \chi 
 - \I \bar \lambda \bar \sigma^m D_m \lambda 
-  {1\over 2}\cD^m \phi \cD_m \phi \nn\\
&& -\ {1 \over 4} \cF^A_{mn} \cF^{Amn} - {1\over4}\cF^B_{mn}\cF^{Bmn}
+ {1\over 2}v^m v_m
  - \Bigl( {1 \over \sqrt{2}}  m \,  \psi^2_m \sigma^m
   \bar \lambda  + m \I \psi^2_m \sigma^m
   \bar \chi 
 \nonumber \\
& &+\ \sqrt{2} m \I \lambda \chi 
 + {1 \over 2} m \chi \chi 
 + \Red m \Black\, \psi^2_m \sigma^{m n} \psi^2_n  
+ {\kappa \over 2\sqrt{2}}  \epsilon_{i j} \psi^i_m \psi_{n}^{j}
    \cF^{Amn}_{-}   \nn\\
    &&+\ {\kappa \over {2}}  \chi \sigma^m
    \bar \sigma^n \psi^1_m \cD_n \phi
+  {\kappa \over 2 }  \bar \lambda
    \bar \sigma_m \psi^1_n \cF^{Bmn}_{+}
    +
  {\kappa \over {2}} 
    \epsilon^{pqrs}
    \bar \psi^2_{p } \bar \sigma_q \psi^1_r
    \cD_s \phi\nn\\
&&- \I\, {\kappa \over {2}}  \chi \sigma^m
    \bar \sigma^n \psi^1_m v_n 
-  \I \, {\kappa \over {2}} 
    \epsilon^{pqrs}
    \bar \psi^2_{p } \bar \sigma_q \psi^1_r
    v_ss
    \plushc \Bigr)  
\eea
where, as before, $m = \kappa v^2$, and
\bea
\cD_m \phi &\ =\ & \partial_m \phi - {m\over\sqrt{2}} (A_m + B_m) \nonumber \\
\cF^A_{mn} &=& \partial_{[m }A_{n]} + {m\over\sqrt{2}} B_{mn} \nonumber \\
\cF^B_{mn} &=& \partial_{[m }B_{n]} - {m\over\sqrt{2}} B_{mn}\ . 
\eea
This Lagrangian is invariant under an ordinary abelian gauge symmetry,
an antisymmetric tensor gauge symmetry, as well as the following two 
supersymmetries,
\bea
\delta_{\eta} e^a_m &\ =\ &\I\, \kappa (\eta^i \sigma^a \overline \psi_{m i} + 
    \bar\eta_i \bar\sigma^a \psi_{m}^i) \nonumber \\
\delta_{\eta} \psi^1_m &=& {2\over\kappa}D_m\eta^1 \nonumber \\
\delta_\eta A_m &=& \sqrt{2}\epsilon_{ij}(\psi_m^i\eta^j + \bar\psi_m^i\bar\eta^j) \nonumber \\
\delta_\eta B_m &=& \bar\eta^1\bar\sigma_m\lambda + \bar\lambda\bar\sigma_m\eta^1\nonumber \\
\delta_\eta B_{mn} &=& 2\eta^1\sigma_{mn}\chi 
+ \I\,\eta^1\sigma_{[ m}\bar\psi^2_{n ]}
+ \I\,\eta^2\sigma_{[ m}\bar\psi^1_{n ]}
\plushc \nonumber\\
\delta_\eta \lambda &=& \I\, \hat\cF^B_{mn}\sigma^{mn}\eta^1 - \I \sqrt{2}
v^2 \eta^2\nonumber\\
\delta_\eta \chi &=& \I\, \sigma^m\bar\eta^1 \hat\cD_m\phi - \hat
v_m\sigma^m\bar\eta^1 + 2 v^2 \eta^2\nonumber \\
\delta_{\eta} \psi^2_m &=& {2\over\kappa}
D_m\eta^2 + \I v^2 \sigma_m \bar\eta^2 
- {\I\over\sqrt{2}}\hat\cF^A_{+mn}\sigma^n\bar\eta^1\nonumber \\
   & & +\ \hat\cD_m \phi \eta^1 +\kappa\left( (\bar\psi^1_m\bar\chi)\eta 
- (\bar\chi\bar\eta)\psi^1_m \right) - \I\, \hat v_m\eta^1\nonumber \\
\delta_\eta \phi &=& \chi\eta^1 + \bar\chi\bar\eta^1 
\eea
up to linear order in the fields.
The supercovariant derivatives are given by
\bea
\hat \cD_m\phi&\ =\ & \cD_m\phi - {\kappa\over{2}}(\psi^1_m\chi +
\bar\psi^1_m\bar\chi)  \nonumber\\
\hat \cF^A_{mn}&=& \cF^A_{mn} + {\kappa\over \sqrt{2}}(\psi^2_{
[m}\psi^1_{n] } + \bar\psi^2_{ [m}\bar\psi^1_{n] }) \nonumber \\
\hat \cF^B_{mn}&=& \cF^B_{mn} - {\kappa\over 2}(\bar\lambda\bar\sigma_{
[n}\psi^1_{m ]} + \bar\psi^1_{ [m}\bar\sigma_{n] }\lambda)  \nonumber\\
\hat v_m &=& v_m + \Bigl(\, \I\kappa \psi^{1}_n\sigma_{m}{}^n\chi  
                 -{\I\kappa\over
2}\epsilon_{m}{}^{nrs}\psi_n^{1}\sigma_r\bar\psi_s^{2} 
\plushc \Bigr)  \ .
\eea
These fields form an irreducible representation of the $N=2$
algebra.

\subsection{The SuperHiggs Effect}

Each of the two Lagrangians presented above has a full $N=2$ supersymmetry
(up to the appropriate order).  The first supersymmetry is realized linearly,
so it is not broken.  The second is realized nonlinearly, so it is
spontaneously broken.  In each case, the transformations imply that
\be
\zeta\ =\ {1\over \sqrt3}\,
(\chi - \I \sqrt 2 \lambda)
\ee
does not shift, while
\be
\nu\ =\ {1\over \sqrt3}\,
(\sqrt 2 \chi + \I \lambda )
\ee
does.  Therefore $\nu$ is the Goldstone fermion for $N=2$ supersymmetry,
spontaneously broken to $N=1$.

In the chiral case, we find
\bea
\left[ \,\delta_{\eta_1}, \,\delta_{\eta_2} \right] \, \phi 
&\ =\ &  2\sqrt{2}\,  v^2\,\eta_1\eta_2  \nonumber \\
\left[ \,\delta_{\eta_1}, \,\delta_{\eta_2} \right] \, {\cal A}_m
&=& {4\over\kappa} \, \partial_m\, \eta_1\eta_2\ .
\eea
The complex scalar $\phi$ is indeed the Goldstone boson
for a gauged central charge.  Moreover, in unitary gauge, where
\be
\phi\ =\ \nu\ =\ 0\ ,
\ee
this Lagrangian reduces to the usual representation for a massive
$N=1$ spin-3/2 multiplet.

In the vector case, we have
\bea
\left[ \, \delta_{\eta^2}, \,  \delta_{\eta^1} \right] \,  A_m &\ =\ &
{2\sqrt{2} \over \kappa}
\partial_m(\eta^1\eta^2 + \bar\eta^1\bar\eta^2) 
  - \sqrt{2} \, \I \,  v^2 \, (\eta^2\sigma_m\bar\eta^1 -
\eta^1\sigma_m\bar\eta^2) \nonumber \\
\left[ \,  \delta_{\eta^2}, \,  \delta_{\eta^1} \right] \,  B_m &=& 
\sqrt{2} \, \I \,  v^2 \, (\eta^2\sigma_m\bar\eta^1 -
\eta^1\sigma_m\bar\eta^2) \nonumber \\
\left[ \,  \delta_{\eta^2}, \,  \delta_{\eta^1} \right] \,  B_{mn} &=&
{2 \, \I\over \kappa}D_{[m }
(\eta^2\sigma_{ n]}\bar\eta^1 - \eta^1\sigma_{n] }\bar\eta^2)\ .
\eea
We see that the real vector $-(A_m - B_m)/\sqrt{2}$ is the 
Goldstone boson for a gauged
{\it vectorial} central extension of the $N=2$ algebra.  In addition,
the real scalar $\phi$ is the Goldstone boson associated with a
single {\it real} gauged central charge.  In unitary gauge, with
\be
-{1\over\sqrt{2}}(A_m - B_m)\ =\ \phi\ =\ \nu\ =\ 0\ ,
\ee
this Lagrangian reduces to the Ogievetsky-Sokatchev
representation for the massive $N=1$ spin-3/2 multiplet.

Now that we have two explicit realizations of partial supersymmetry
breaking, we can go
back and see how they avoid the no-go argument presented in the
introduction.  We first compute the second supercurrent.  In each
case it turns out to be
\be
J^2_{m\alpha}\ =\ 
v^2\,(\sqrt6\, \I \,\sigma_{\alpha\dot\alpha m}
\bar\nu^{\dot\alpha} +
4\,\sigma_{\alpha\beta m n} \psi^{2n\beta})
\ee
plus higher-order terms.  Computing, we find
\bea
\{\,\bar Q_{\dot\alpha},\,J^1_{m\alpha}
\,\} &\ =\ & 2\,\sigma^n_{\alpha\dot\alpha}\,
T_{mn}\nonumber \\
\{\,\bar S_{\dot\alpha},\,J^2_{m\alpha}
\,\} &=& 2\,\sigma^n_{\alpha\dot\alpha}\,
T_{mn}\ .
\eea
In the presence of supergravity, there is no confusion
about the stress-energy tensor.  There is just one
such tensor, and it shows up on the right-hand side of the current
algebra.

For the case at hand, however, $J^i_{\alpha m}$ and $T_{mn}$ contain
contributions
from {\it all} of the fields, including the second gravitino.  When
covariantly-quantized, the second gravitino gives rise to states of
negative norm.  Indeed, it is not hard to check that
\be
(\bar S S + S \bar S)\,
|0\rangle\ =\ 0\ ,
\ee
even though
\be
S\,|0\rangle\ \ne\ 0 \quad\qquad
\bar S\,|0\rangle\ \ne\ 0\ .
\ee
The supergravity couplings exploit the second loophole to the
no-go theorem!

The Lagrangian in the chiral case is a truncation of the supergravity
coupling found by Cecotti, Girardello and Porrati (1986) and by
Zinov'ev (1987).  Their results were based on {\it linear} $N=2$
supersymmetry; they involved $N=2$ vector- and hyper-multiplets.  The
Lagrangian for the vector case is new.  It contains a new realization
of $N=2$ supergravity.  In each case, the couplings presented here
are minimal and model-independent.  They describe the superHiggs
effect in the low-energy effective theories that arise from partial
supersymmetry breaking.

Thus we have seen that there is no obstacle to partial supersymmetry
breaking in the presence of gravity.  Indeed, each of the two
Goldstone multiplets give rise to its own massive spin-3/2 multiplet.
Of course, the connection between these results and the theory of
membranes and D-branes is an urgent and open question.

\vspace{0.5in}

It is a pleasure to thank Sam Osofsky and Richard
Altendorfer for collaboration on the supergravity couplings
presented here.
We would also like to acknowledge our debt to Victor Isaacovitch
Ogievetsky for teaching us the physics that underlies this work,
and so much more.  This work was supported by the National
Science Foundation, grant NSF-PHY-9404057.


\begin{thebibliography}
%
\bibitem{}{ABO}{}
Altendorfer,\,R., Bagger,\,J., Osofsky,\,S. (1998):
In preparation. \\
See also Altendorfer,\,R., Bagger,\,J. (1998):
New Supersymmetry Algebras from Partial Supersymmetry Breaking.
hep-th/9809171.
%
\bibitem{}{APT}{}
Antoniadis,\,I., Partouche,\,H., Taylor,\,T. (1996):
Spontaneous Breaking of $N=2$ Global Supersymmetry.
Phys.\,Lett. {\bf B372}, 83.
%
\bibitem{}{bg1}{}
Bagger,\,J., Galperin,\,A. (1994):
Matter Couplings in Partially Broken Extended Supersymmetry.
Phys.\,Lett. {\bf B336}, 25.
%
\bibitem{}{bg2}{}
Bagger,\,J., Galperin,\,A. (1997a):
New Goldstone Multiplet for Partially Broken Supersymmetry.
Phys.\,Rev. {\bf D55}, 1091.
%
\bibitem{}{bg3}{}
Bagger,\,J., Galperin,\,A. (1997b):
The Tensor Goldstone Multiplet for Partially Broken Supersymmetry.
Phys.\,Lett. {\bf B412}, 296.
%
\bibitem{}{cgp}{}
Cecotti,\,S., Girardello,\,L., Porrati,\,M. (1986):
An Exceptional $N=2$ Supergravity with Flat Potential and Partial SuperHiggs.
Phys.\,Lett. {\bf 168B}, 83.
%
\bibitem{}{cwz}{}
Coleman,\,S., Wess,\,J., Zumino,\,B. (1969):
Structure of Phenomenological Lagrangians. 1.
Phys.\,Rev. {\bf 177}, 2239.
%
\bibitem{}{FV}{}
Ferrara,\,S., van Nieuwenhuizen,\,P. (1983):
Noether Coupling of Massive Gravitinos to $N=1$ Supergravity.
Phys.\,Lett. {\bf B127}, 70.
%
\bibitem{}{HLP}{}
Hughes,\,J., Liu,\,J., Polchinski,\,J. (1986):
Supermembranes.
Phys.\,Lett. {\bf 180B}, 370.  \\
See also Hughes,\,J.,  Polchinski,\,J. (1986):
Partially Broken Global Supersymmetry and the Superstring.
Nucl.\,Phys. {\bf B278}, 147.
%
\bibitem{}{inverse}{}
Ivanov,\,E., Ogievetsky,\,V. (1975):
The Inverse Higgs Phenomenon in Nonlinear Realizations.
Teor.\,Mat.\,Fiz. {\bf 25}, 164.
%
\bibitem{}{gaugessb}{}
Ivanov,\,E., Ogievetsky,\,V. (1976):
Gauge Theories as Theories of Spontaneous Breakdown.
JETP\,Lett. {\bf 23}, 606.
%
\bibitem{}{ogsok}{}
Ogievetsky,\,V., Sokatchev,\,E. (1976):
On Gauge Spinor Superfield.
JETP Lett. {\bf 23}, 58. \\
See also Gates,\,S.J. (1977):
Spinor Yang-Mills Superfields.
Phys.\,Rev. {\bf D16}, 1727.
%
\bibitem{}{OP}{} 
Ogievetsky,\,V., Polubarinov,\,I. (1966):
The Notoph and Its Possible Interactions.
Sov.\,J.\,Nucl.\,Phys. {\bf 4}, 156.
%
\bibitem{}{volkov}{}
Volkov,\,D.V. (1973):
Phenomenological Lagrangians.
Sov.\,J.\,Particles and Nuclei {\bf 4}, 3.
See also Ogievetsky,\,V.I. (1974):
Nonlinear Realizations of Internal and Spacetime Symmetries,
in {\it Proceedings of the X-th Winter School of Theoretical
Physics in Karpacz,} (Wroclaw), 227.
%
\bibitem{}{zin}{}
Zinov'ev,\,Yu M. (1987):
Spontaneous Symmetry Breaking in $N=2$ Supergravity.
Sov.\,J.\,Nucl.\,Phys. {\bf 46}, 540.
%
\end{thebibliography}
\end{document}